\documentclass[twocolumn,prd]{revtex4} 
\usepackage{graphicx}

\begin{document}


\title{On the large-scale instability in interacting dark energy and
  dark matter fluids}


\author{Brendan M Jackson}
\email{bmj@roe.ac.uk}
\author{Andy Taylor}
\affiliation{Institute for Astronomy, School of Physics and
Astronomy, University of Edinburgh, Royal Observatory, Blackford Hill, Edinburgh EH9 3HJ}
\author{Arjun Berera}
\affiliation{Institute for Physics, School of Physics and Astronomy,
  University of Edinburgh, James Clerk Maxwell Building, The King's
  Buildings, Mayfield Road, Edinburgh EH9 3JZ}


\date{\today}

\begin{abstract}

Recently, Valiviita et al. (2008) have reported a
large-scale early-time instability in coupled dark energy and dark
matter models. We take the same form of energy-momentum exchange and
specialise to the case when the interaction rate is proportional to
Hubble's parameter and the dark energy density only. Provided the
coupling is made small enough for a given equation of state parameter,
we show that the instability can be avoided. Expressions are derived
for non-adiabatic modes on super-horizon scales in both the radiation
and matter dominated regimes. We also examine the growth of dark
matter perturbations in the sub-horizon limit. There we find that the
coupling has almost no effect upon the growth of structure before dark
energy begins to dominate. Once the universe begins to accelerate, the
relative dark matter density fluctuations not only cease to grow as in
uncoupled models, but actually decay as the universe continues to
expand.

\end{abstract}

\pacs{}

\maketitle

\section{Introduction}

There is good evidence to believe the present day energy density of
the universe is mostly in the form of dark energy
\cite{2008WMAP,2005BAO}, the properties of which remain relatively
unknown. Furthermore, observations of Type Ia supernovae 
\cite{1999Perl,1998Riess} leave little doubt that the expansion of the 
universe is accelerating. Viable models of cosmology now require a
large dark energy component, capable of producing the negative
pressure required for accelerated expansion. 

By far the simplest model of dark energy is Einstein's cosmological
constant, $\Lambda$. The cosmological constant($\Lambda$) and cold
dark matter (CDM) model, with values of today's density parameter for
the dark energy $\Omega_x \approx 0.7$ and dark matter $\Omega_m
\approx 0.3$ is the current prevailing paradigm. But while consistent
with observational constraints, the standard model is in many ways
unsatisfactory. One such example is the `coincidence problem': why are
the energy densities in the dark energy and dark matter comparable
today, when the redshift dependence of each is so different?

Motivated to explain the coincidence problem while deviating as little
as possible from the successful $\Lambda$CDM model, a coupling between
dark energy and dark matter has often been considered. An energy
exchange modifies the background evolution of the dark sector, and
explaining the coincidence problem can be reduced to tuning a coupling
parameter to an appropriate value. 

The coupling enters via the continuity equations. \nolinebreak
With energy exchange rate $Q$ between the dark energy
(subscript $x$) and the cold dark matter (subscript $c$), the
dark energy obeys the continuity equation in conformal time
\begin{equation}
\dot{\rho_x} + 3\mathcal{H}(1+w_x)\rho_x = -Q,
\end{equation}
while the dark matter obeys
\begin{equation}
\dot{\rho_c} + 3\mathcal{H}\rho_c = Q.
\end{equation}
Here we have introduced the equation of state parameter $w_A$ that gives
the ratio of the pressure $P_A$ to the energy density $\rho_A$ of a
fluid,
\begin{equation}
w_A = \frac{P_A}{\rho_A}.
\end{equation}
We have also used $\mathcal{H} = aH$, where $a(t)$ is the expansion
scale-factor and $H$ the Hubble parameter. Acceleration of the
expansion rate requires the energy density of the universe to
dominated by a fluid with an effective equation of state parameter
$w_{\rm{eff}} < -1/3$. We do not allow the phantom case of $w < -1$ in
this work.  Simple solutions for the background exist for couplings of
the form $Q = \alpha \mathcal{H} \rho_x +\beta \mathcal{H}
\rho_c$. These were initially investigated by Chimento \cite{1997Ch}
and then expanded upon by Barrow and Clifton \cite{2006B}, who
provided general solutions for any cosmology with two components
exchanging energy in such a fashion, provided the components were
modelled as cosmological fluids with constant $w$.  Quartin et
al. \cite{2008Q} examined the observational constraints upon such a
class of models, significantly limiting the available parameter space.
Again, the equation of state parameter was treated as fixed. Non-zero
values of $\beta$ were found to reduce the required fine-tuning of the
initial energy density, as well as increase the observationally
allowed values of $\alpha$ \cite{2008Q}.

A coupling would influence more than just the background dynamics of
the universe. In particular, the growth of perturbations in the
coupled fluids would be affected. Recent work by Valiviita et
al. \cite{Valiviita2008} has shown that couplings of the simple form
described above, with constant $w$, exhibit extremely rapid growth of
dark energy fluctuations on super-horizon scales in the early
universe. In fact, the perturbations in the dark energy become
unstable for any model with non-zero $\beta$, no matter how small this
parameter is made. While this would appear to rule out all couplings
of the above form and with constant $w$, the explicit examples in
\cite{Valiviita2008} included no cases where the interaction rate was
proportional to the density of dark energy and not of the dark
matter, i.e. with $\beta = 0$ and $\alpha \neq 0$. Here we look at 
just such a scenario. 


\section{Background evolution}\label{background}

Friedmann's equation relates the  evolution of the scale-factor $a(t)$
to the  background energy  density $\rho$. We  make use  of conformal
time,  $\tau$,  which   is  related  to  cosmic  time   via  $dt  =  a
d\tau$.  Overdots  indicate  derivatives  with  respect  to  conformal
time. 

Friedmann's equation reads
\begin{equation}
\mathcal{H}^2 \equiv \left(\frac{a^{\prime}}{a} \right)^2 = \frac{8 \pi
G}{3} \rho a^2.
\end{equation}

With the choice of $Q = \alpha \mathcal{H} \rho_x$, the continuity
equations can be solved to yield \cite{2006B,1997Ch}
\begin{equation}
\rho_x = \rho_{x,0} \; a^{-(3(1+w)+\alpha)},
\end{equation}
\begin{eqnarray}
\rho_c &=& -\frac{\alpha \rho_{x,0}}{3w+\alpha} \; a^{-(3(1+w)
         +\alpha)} \nonumber \\ &+& \left(\rho_{c,0} + \frac{\alpha
         \rho_{x,0}}{3w+\alpha}\right) \; a^{-3}.
\end{eqnarray}
We follow the standard notation where a subscript zero indicates
today's value. We normalise the scale-factor so that $a_0 = 1$.
The ratio of dark energy to dark matter density $r$ can then be
written
\begin{equation}
\frac{1}{r} = \frac{\rho_{c}}{\rho_{x}} =  
              \left( \frac{\rho_{c,0}}{\rho_{x,0}} + 
              \frac{\alpha}{3w+\alpha} \right) a^{3w+\alpha}
              - \frac{\alpha}{3w+\alpha}
\end{equation}
With $|3w| < \alpha$, the dark energy and dark matter approach a
constant ratio as the universe expands. The coincidence problem can be
said to be solved if this ratio is of order unity, but this requires a
value of $\alpha$ already observationally excluded \cite{2008Q}. 
Nevertheless, as argued in \cite{2008Q}, non-zero values of $\alpha$
can still be said to alleviate the problem. We restrict ourselves to 
positive values of $\alpha$.
 
\section{Perturbed FRW Cosmology} \label{perturbed}

We assume a flat FRW cosmology and work in Newtonian gauge, 
\begin{equation}
-ds^2  = dt^2(1+2\Psi) - a^2 (1-2\Phi)\delta_{i j}dx^i dx^j,
\end{equation}
with metric signature $(+,-,-,-)$. We work in Fourier space, using comoving
Fourier wave-vectors $k^i = k_i$, so that $\partial_i \partial^i \to k^2/a^2$.

The four-velocity of fluid $A$ is given by
\begin{equation}
U^\mu_{(A)} =  \left((1-\Psi), a^{-1} v^i_{(A)} \right).
\end{equation}
The peculiar velocity three-vector $v^i=v_i$ are small.
We define the velocity perturbation $\theta \equiv \partial_i v^i$.

\subsection{Energy-momentum tensors}

The energy-momentum tensor for fluid $A$ is given by:
\begin{equation}
T^{\mu (A)}_{\; \; \nu} = \left( \rho^{(A)} + P^{(A)} \right) U^{(A)\mu} U^{(A)}_\nu -
\delta^\mu_{\,\nu} P^{(A)}.
\end{equation}
The total energy-momentum tensor is simply the sum of the components,
\begin{equation}
T^{\mu}_{\; \; \nu} = \sum_A T^{\mu (A)}_{\; \; \nu}.
\end{equation}

We define the density perturbation in fluid $\delta_A$ using
$\rho_A \equiv (1+\delta_A)\bar{\rho}_A$. An overbar denotes the 
background quantity, though we will usually leave this implicit.

Energy and momentum conservation for fluid $A$ implies
\begin{equation}
\nabla_{\mu} T^{\mu (A)}_{\; \; \nu} = Q^{(A)}_\nu, \nonumber
\end{equation}
and conservation for the entire system requires
\begin{equation} 
\sum_A Q^\mu _{(A)} = 0.
\end{equation}
The four-vector $Q^\mu _{(A)}$ governs the energy exchange between
components, and it is to this we now turn our attention.

\subsection{Covariant energy exchange}

The energy exchange in the background does not determine a fully
covariant form of energy exchange
\cite{K2005,Valiviita2008}. Instead, an energy exchange four-vector
must be specified. We adopt the approach of \cite{Valiviita2008} and
consider two scenarios; aligning the four-vector with the dark energy
four-velocity,
\begin{equation}\label{dark energy align}
Q^\mu_{(A)} = Q_A U^\mu_x,
\end{equation}
or with the four-vector of the dark matter four-velocity,
\begin{equation}\label{dark matter align}
Q^\mu_{(A)} = Q_A U^\mu_c.
\end{equation}
These choices produce slightly different outcomes, and the differences are
noted as we proceed.

To produce the desired changes to the continuity equations, we see
that $aQ_c = -aQ_x = \alpha \mathcal{H} \rho_x$ in both cases. We also
make the common assumption that $\alpha \mathcal{H}$ gives an
interaction rate that has no spatial dependence. We therefore perturb
only $\rho_x$, not $\mathcal{H}$, in the coupling.

\subsection{Sound speed of dark energy}

The speed of sound of a fluid or scalar field $A$ is denoted by
$c_{s A} $. For a barotropic fluid with a constant value of $w_A$, 
then $c_{s A}^2 = w_A$. This leads to an imaginary speed of sound for
the dark energy ($c_{s x}^2 = w_x < 0$). An imaginary sound speed leads
to instabilities in the dark energy; the problem is commonly remedied
by imposing a real sound speed by hand. A common choice (and the one we make 
here) is the scalar field value of $c_{s x} = 1$. 

This choice leads to an intrinsic non-adiabatic pressure perturbation
in the dark energy. This contains a term, highlighted recently in
\cite{Valiviita2008}, that arises due to the coupling between dark
energy and dark matter. We include this term, and refer the interested
reader to \cite{Valiviita2008}.

\subsection{Perturbation equations of motion}

Conservation of the energy-momentum tensor, combined with results of
the previous sections and our choice of energy exchange four-vector,
implies the following.
For the dark energy density perturbation:
\begin{eqnarray} \label{delta x}
\delta_x^\prime &+& 3\mathcal{H} (1-w_x) \delta_x + (1+w_x) \theta_x +
9\mathcal{H}^2 (1-w_x^2)\frac{\theta_x}{k^2} \nonumber \\&-&
3(1+w_x)\Phi^\prime= -\alpha \mathcal{H} \left[\Psi +
3\mathcal{H}(1-w_x) \frac{\theta_x}{k^2} \right].
\end{eqnarray}
For the dark energy velocity perturbation, the right-hand side differs slightly depending
on our choice of energy exchange four-vector.
\begin{samepage}
\begin{eqnarray}\label{theta x}
\theta_x^\prime - 2\mathcal{H}\theta_x &-& \frac{k^2}{1+w_x} \delta_x - k^2\Psi \nonumber \\ &=& 
\begin{array}{ll}
{\displaystyle \frac{(1+b)\alpha \mathcal{H}}{1+w_x} \theta_x},
\end{array}
\end{eqnarray}
\end{samepage}
where
\begin{equation}
b = \left\{
\begin{array}{rl}
0 & \text{if } Q^\mu_{(A)} = Q_A U^\mu_x, \\
1 & \text{if } Q^\mu_{(A)} = Q_A U^\mu_c.
\end{array}
\right.
\end{equation}

For the dark matter, the density perturbation obeys
\begin{equation} \label{delta c}
\delta_c ^\prime + \theta_c -3\Phi^{\prime} = 
\alpha \mathcal{H}\frac{\rho_x}{\rho_c}\left[ \delta_x-\delta_c
  \right],
\end{equation}
while the velocity perturbation is governed by
\begin{eqnarray} \label{theta c}
\theta^\prime_c &+& \theta_c \mathcal{H} - k^2 \Psi \nonumber \\ &=& 
\begin{array}{ll} 
{ \displaystyle (1-b)\alpha\mathcal{H} \frac{\rho_x}{\rho_c} \left[\theta_x - \theta_c \right]}. \\
\end{array}.
\end{eqnarray}

The perturbed Einstein equations are well known, and we do not
reproduce them here. They can be found in \cite{1995Ma}, whose
notation for the scalar metric perturbations we share. 

\section{Initial conditions in the early radiation era} \label{early}

In \cite{Valiviita2008} it was shown that models with $\beta \neq 0$
suffered from an early time large-scale instability no matter how
small the value of $\beta$.  This was driven by a term proportional to
$\beta$ on the right-hand side of equation (\ref{theta x}).  A term
proportional to $\alpha$ also exists, which can be large if $w$ is
close to $-1$ or $\alpha$ is made very large. In this section we
examine how large this term needs to be to cause the non-adiabatic
mode to be a growing one.

We consider super-horizon scales ($k/\mathcal{H} \ll 1$) and assume
adiabatic initial conditions. The gravitational potentials are
dominated by fluctuations in the dominant fluid (radiation or
matter). The well known result is that $\Phi \propto \Psi =
\text{constant}$. The constant of proportionality in the radiation era
is determined by the anisotropic stress generated by the neutrinos. In
the absence of neutrinos or in the matter dominated era, the
potentials are equal. These assumptions will be invalid only if
perturbations in the dark energy are large enough to influence the
gravitational potentials. As the dark energy has a very low background
density in the radiation era, this can only happen if $\delta_x$ grows
extremely large.

Neglecting time derivatives of the gravitational potential, and
keeping only leading order terms in $k/\mathcal{H}$, the dark energy
equations (\ref{delta x}) -- (\ref{theta x}) can be combined into a
second order equation:
\begin{eqnarray}\label{second order de}
\delta_x^{\prime \prime} &+& \mathcal{H}
\left(1-3w-\frac{(1+b)\alpha}{1+w} -
2\frac{\mathcal{H}^\prime}{\mathcal{H}^2} \right)\delta_x^{\prime}
\nonumber \\ &+& 3\left(\mathcal{H}^2 \left(1-b\frac{\alpha}{1+w}
\right) - \mathcal{H}^\prime \right)(1-w) \, \delta_x \nonumber \\ &=&
\left(A\mathcal{H}^2 + B\mathcal{H}^\prime \right) \Psi,
\end{eqnarray}
The constants $A$ and $B$ have values unimportant for our analysis.

In the radiation era, $\mathcal{H} = \tau^{-1}$. The adiabatic mode is
therefore an obvious solution: $\delta_x \propto \Psi =
\text{constant}$. To find the remaining solutions, we define a new
variable $\hat{\delta}_x = \delta_x + C\Psi$, with the constant $C$
chosen such that the right-hand side of (\ref{second order de}) is
equal to zero. In the radiation dominated era, we can then write:
\begin{eqnarray}\label{final early time second order}
\tau^2 \hat{\delta}_x^{\prime \prime} &+&
\left(3-3w-\frac{(1+b)\alpha}{1+w} \right)\tau \hat{\delta}_x^{\prime} 
\nonumber \\
&+& 3(1-w)\left(2-b\frac{\alpha}{1+w}\right)\hat{\delta}_x = 0
\end{eqnarray}
When $b=1$, equation (\ref{final early time second order}) becomes
formally the same equation found by He et al.\cite{He:2008}, despite
the differing assumptions made about the physics involved.  In their
investigation of perturbations given a background coupling of the form
$Q = \alpha\mathcal{H}\rho_x$, they choose to set the net momentum
exchange to zero ($Q^i_{(A)} = 0$), in contrast to our adoption of the
form of momentum exchange used in \cite{Valiviita2008}. The
differences between the $b=1$ choice of momentum exchange and zero net
momentum exchange arise in the equations for the dark matter
perturbations, which are not used in the above analysis, nor in the
analysis by He et al. This leads to the same behaviour of dark energy
perturbations. This is not true when $b=0$, and can result in
different behaviour (oscillatory or non-oscillatory) for the same
choice of parameters (see the remainder of this section). Note also
that the simplifying assumptions, and their justifications, made in
\cite{He:2008} differ to those made here: we have neglected terms that
will be small due to choice of intial conditions, and simplified the
result by extracting the adiabatic mode. In \cite{He:2008}, terms are
instead neglected that are found to be small from a numerical
analysis.

Solutions of equation (\ref{final early time second order}) are power
laws, $\hat{\delta}_x \propto \tau^{n_{\pm}}$. The index is given by:
\begin{equation}
n_{\pm} = \frac{\Gamma}{2(1+w)} \pm \frac{\sqrt{\Delta}}{2(1+w)},
\end{equation}
where we follow the notation of \cite{He:2008} and have defined
the quantities
\begin{equation}
\Gamma = 3w^2 + w + (1+b)\alpha - 2,
\end{equation}
and
\begin{eqnarray}
\Delta &=& 9w^4 + 30w^3 + (13-6(b-1)\alpha)w^2 \nonumber \\ &+&  
 2w\left[(1+b)\alpha -14\right] +  4(2b-1)\alpha  
\nonumber \\ &+& (1+b)^2 \alpha^2 - 20.
\end{eqnarray} 

In the limit of $w$ very close to -1 (and assuming $\alpha$ is
reasonably small), we can expand as a series in $(1+w)$,
\begin{eqnarray}
\frac{\Gamma}{2(1+w)} \approx -5/2 + \frac{(1+b)\alpha}{2(1+w)} &+&
\frac{3(1+w)}{2} \nonumber \\ &+& \mathcal{O}(1+w)^2,
\end{eqnarray}
\begin{eqnarray}
\Delta &\approx& (1+3b)\alpha^2 + 2(7b-5)(1+w) \alpha  
\nonumber \\
&+& (6(1-b)\alpha -23)(1+w)^2 + \mathcal{O}(1+w)^3.
\end{eqnarray}
When $\alpha = 0$, the non-adiabatic mode is decaying. But when the
coupling is switched on, the second term in $\Gamma$ can become very
large, resulting in $n_+ \gg 1$. For a range of $\alpha$ and $w$,
which is much larger in the  $b=0$ case, oscillatory behaviour can also
result (due to $\Delta$ becoming negative).  The instability means
these coupled models suffer from all the problems outlined in
\cite{Valiviita2008} for $\beta \neq 0$ models, unless the value of
$\alpha$ is made small enough. The closer $w$ is to $-1$, the smaller
$\alpha$ must be made to avoid the instability. This is in contrast to
$\beta \neq 0$ models, which are unstable no matter how small the
parameter $\beta$ is made.

In the matter dominated era, we can carry out the same procedure, this
time with $\mathcal{H} = 2\tau^{-1}$. We find,
\begin{equation}
\frac{\Gamma}{2(1+w)} \approx -9/2 + \frac{(1+b)\alpha}{1+w} + 3(1+w)
+ \mathcal{O}(1+w)^2,
\end{equation}
and
\begin{eqnarray}
\Delta &\approx&  4(1+3b)\alpha^2 + 12(5b-3)\alpha(1+w) \nonumber \\
&+& (24(1-b)\alpha-62)(1+w)^2 + \mathcal{O}(1+w)^3.
\end{eqnarray}
Once again, the second term in $\Gamma$ can result in a rapidly
growing dark energy fluctuation.

\begin{figure}
   \includegraphics[totalheight=0.25\textheight,viewport=40 29 940 653,clip,angle=0]{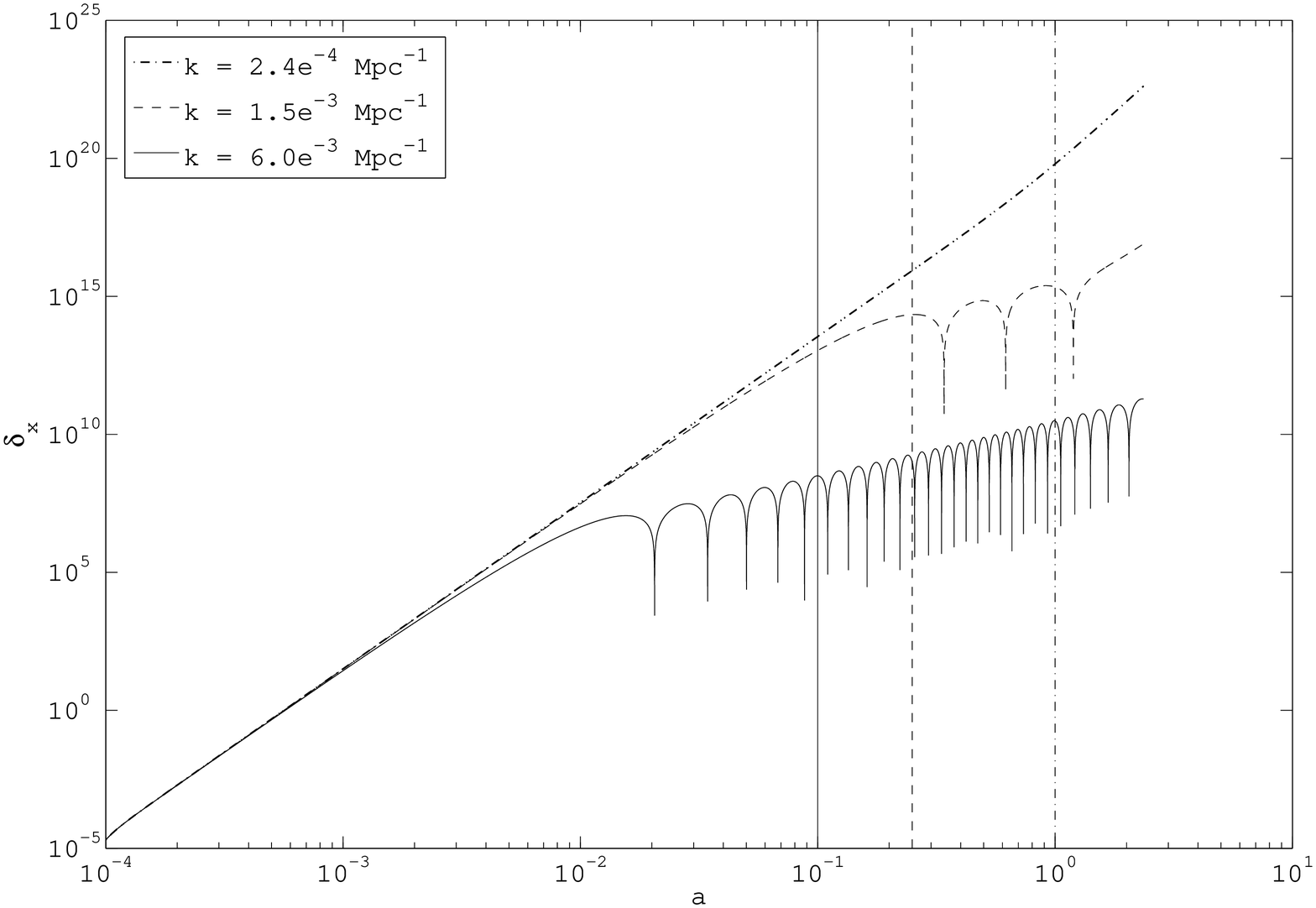}
\caption{\label{graph1}The evolution of the dark energy outside the
  horizon in a matter dominated universe, for modes of three
  different scales. We take $\alpha = 0.08$ and $w=-0.98$, and
  $b=1$. The agreement with the analytical approximation is excellent
  until the mode begins to leave the horizon ($k\tau \sim
  1$). Vertical lines indicate when this occurs for each mode.}
\end{figure}

We have solved equations (\ref{delta x}) -- (\ref{theta c})
numerically in the matter dominated regime (Figure \ref{graph1}),
where we need not worry about the radiation fluid and its
perturbations. The analytical agreement is excellent until the mode
leaves the horizon ($k\tau \sim1$). Numerically we see that when this
happens the mode begins to oscillate with a growing amplitude.

\section{Sub-horizon evolution in the matter and radiation dominated eras}\label{sub}

In the sub-horizon limit, $\mathcal{H}^2/k^2 \ll 1$, equation
(\ref{delta x})  yields,
\begin{equation} \label{small delta x}
\delta_x^\prime + 3\mathcal{H}(1-w_x)\delta_x + (1+w_x) \theta_x
-3(1+w_x)\Phi^\prime = 0.
\end{equation}
Note the two terms on the right-hand side of equation (\ref{delta x})
scale as $\mathcal{H}^2/k^2$. As these are the only two terms
containing the coupling parameter $\alpha$, the simplified equation
above does not contain the coupling parameter.

One of the perturbed Einstein equations simplifies to Poisson's
equation in comoving coordinates,
\begin{equation}
-k^2\Psi = 4\pi G a^2 \left(\rho_x \delta_x + \rho_c \delta_c \right).
\end{equation}
Without the coupling, the dark energy perturbations are significantly
suppressed on small-scales in comparison to dark matter perturbations,
primarily due to its large speed of sound \cite{Bean}. The coupling
does nothing to alter this fact unless the right-hand side of equation
(\ref{theta x}) makes a significant contribution. If the early time
instability has been avoided this cannot be the case, as
$\displaystyle \alpha/(1+w)$ will be small. Thus it is reasonable to expect
the dark energy to remain suppressed on sub-horizon scales. We
therefore neglect dark energy perturbations for the remainder of this
section.

By combining equations (\ref{delta c}) and (\ref{theta c}), we eliminate
$\theta_c$ and find a second-order equation for the growth of
the matter density perturbation. From the above argument, we have
neglected dark energy perturbations. Keeping only the dominant
gravitational terms,
\begin{samepage}
\begin{eqnarray} \label{second order w}
\delta_c^{\prime \prime} &+& \mathcal{H}\left(1+2\alpha
\frac{\rho_x}{\rho_c} \right) \delta_c^\prime \nonumber \\
&+& \alpha \frac{\rho_x}{\rho_c}\left(\mathcal{H}^\prime -
\mathcal{H}^2(\alpha + 3w - 1) \right)\delta_c = -k^2 \Psi.
\end{eqnarray}
\end{samepage}
We note that in the limit of $\alpha \to 0$, this reduces to the
standard growth equation, with the well known growing mode $\delta_c
\propto \tau^2$ in both matter and radiation eras. The additional terms are 
proportional to $\alpha r$ (recall $r$ is the ratio of dark energy to dark matter).  
In the matter dominated regime, then $\alpha r \ll 1$, and these terms will be
negligible. Even when $r \sim 1$, the terms will be suppressed by the
size of $\alpha$, which will be small itself. The dominant effect
causing a deviation from standard linear growth of structure in the
matter dominated regime will therefore be, as in an uncoupled
cosmology, the influence of the dark energy upon the expansion
rate. The growth of structure in a coupled model can therefore be
treated in the matter dominated regime simply as an uncoupled model
with an effective dark energy equation of state parameter
$\displaystyle w_{\rm{eff}} = w + \alpha/3$. This will cease to be
true only when the background energy density of matter is no longer
well approximated by its usual $\rho_c \propto a^{-3}$ dependence, and
the late time scaling behaviour becomes apparent.

\section{Sub-horizon evolution in the dark energy dominated era}\label{late}

The coupling between dark energy and dark matter eventually leads to a
constant ratio between the two dark components. With a small value of
$\alpha$, the dark energy still dominates. We consider the evolution
of structure once this equilibrium has been reached. 

Friedmann's equation solves to yield 
\begin{equation}
\mathcal{H}=2(\alpha+3w+1)^{-1}\eta^{-1},
\end{equation}
with the new time variable $\eta = \tau - \tau_\infty$.  Note that as
$\eta$ increases ($\tau$ decreases and approaches $\tau_\infty$), the
scale-factor increases.  The constant of integration, $\tau_\infty$,
is the radius of the de Sitter event-horizon in the uncoupled case
with a cosmological constant.  The growth equation can then be written
as,
\begin{samepage}
\begin{eqnarray}
  \eta^2 \delta_c^{\prime \prime} &+& \eta \frac{2-12w-4\alpha}{\alpha+3w+1} 
\delta_c^{\prime} \nonumber \\ &+& 2\frac{(3\alpha +9w - 1)(3w+\alpha) +
\alpha w}{(\alpha+3w+1)^2}\delta_c = 0.
\end{eqnarray}
\end{samepage}

\begin{samepage}
This admits power law solutions, $\delta_c \propto \eta^m$, where
\begin{equation}
m = \frac{5}{2} - \frac{3}{1+3w+\alpha}  \pm \frac{1}{2} \sqrt{1-\frac{8\alpha}{w(1+3w+\alpha)^2}}.
\end{equation}
\end{samepage}
In the range of $\alpha$ and $w$ relevant to the problem, then $m>0$.
Recalling that $\eta$ decreases with increasing scale-factor, we see
that the universe becomes steadily more homogeneous as it
expands. We interpret this to be a combination of two effects. The
first is the accelerating expansion, which slows and (without the
coupling) eventually stops structure formation. This occurs, for
example, in $\Lambda$CDM cosmology when the cosmological constant
becomes dominant. The second effect is that dark energy is constantly being 
transformed into dark matter, via the coupling. As the rate is proportional 
to the density of the dark energy, and the dark energy density is
essentially uniform, new dark matter is also created uniformally. This
rising `background' of dark matter reduces the relative value of the
fluctuations, reducing $\delta_c$.

We have also investigated both numerically and analytically the
extreme late time behaviour, where $k\eta \ll 1$ and the modes can be
thought of as leaving the horizon. We find the tend toward homogeneity
continues, but with a much milder rate of decay.

\section{Conclusions}\label{conclusions}

We have shown that constant $w$ models with the same form of
energy-momentum exchange considered by \cite{Valiviita2008} suffer
from an instability with $\alpha \neq 0$, even if $\beta = 0$.
However the instabilities in these models are not as severe as those
facing models with $\beta \neq 0$. There is at least some non-trivial
region of parameter space where the instability can be avoided,
although the value of $\alpha$ is now constrained both from background
observables \cite{2008Q} and from stability requirements to be
extremely small. Despite this, any non-zero value of $\alpha$ will
lead to a late-time scaling regime, alleviating (even if not solving)
the coincidence problem. It is unfortunate that with $\alpha$
constrained to such small values, we find any observable trace upon
the growth of CDM structure will be negligible. Detecting a coupling
of this form from measurements of large-scale structure is extremely
doubtful, even with the precision promised by future experiments.

We have said nothing up to this point of models of dark energy with a
variable equation of state parameter, such as scalar-field
(quintessence) models. The same caveats in \cite{Valiviita2008} apply
here. Much of the above analysis will not apply in variable $w$
models, although some parameterisations such as the often used 
$w = w_0 + (1-a) w_a$ lead to fixed $w$ over large periods 
of time. Our analysis will apply during those epochs of constant $w$. 
We refer interested readers to recent work on quintessence with
couplings of this or similar form (such as recent work 
\cite{Chong2008,Corasaniti2008}).
 
The future decay of dark matter fluctuations is an interesting result.
It implies observers today find themselves close to the time of
maximum inhomogeneity.  The more the coincidence problem is
alleviated, the closer to the late time scaling regime today becomes,
and thus the closer to the peak of inhomogeneity.  Without the
coupling, observers find themselves at the time of the end of
structure growth.  The root cause in both cases is the acceleration of
the universe only beginning today. Our position as apparently
privileged observers in this fashion remains difficult to explain in
any satisfactory way.

\begin{acknowledgments}

B.M.J. acknowledges the support from theSTFC. We also thank Fergus
Simpson for bringin interaction models to our attention.

\emph{Note added in proof:} after submission we became aware of the
work by \cite{Gavela2009} in which an independant analysis of general
instabilities in coupled models was carried out, as well as more
specifically the model considered in this work. They provide
cosmological parameter constraints for negative values of
$\alpha$ and find a non-trivial region consistent with observations.

\end{acknowledgments}

\bibliography{refs}

\end{document}